\documentclass[dvips, a4paper, 12pt]{article}

\usepackage{indentfirst}

\usepackage[english]{babel}
\usepackage[latin1]{inputenc}

\usepackage{color}
\definecolor{oneblue}{rgb}{0,0.0,0.75}
\usepackage[colorlinks,
            urlcolor=oneblue,
            linkcolor=oneblue,
            citecolor=oneblue,
            bookmarksopen=false,
            pdfpagemode=FullScreen,
            pagebackref]{hyperref}

\hypersetup{
    pdftitle = Dynamics of tsunami waves,
    pdfauthor = Frédéric Dias andDenys Dutykh,
    pdfsubject = Tsunami waves,
    pdfkeywords = {Duynamics of tsunami, Boussinesq equations, long waves approximation, tsunami propagation}
}

\usepackage[dvips,final]{graphicx}

\usepackage[sort,sectionbib]{natbib}

\usepackage{xspace} 
\usepackage{amsmath,amssymb,amsthm}

\usepackage{fancyhdr}
\def\rD{{\rm D}}
\def\uv{\mathbf{u}}

\def\Dv{\mathbf{D}}

\def\sech{\mbox{sech}}
\def\fr{\hbox{$\frac{1}{2}$}}
\def\onet{\hbox{$\frac{1}{3}$}}
\def\ones{\hbox{$\frac{1}{6}$}}
\def\onetf{\hbox{$\frac{1}{24}$}}

\newcommand{\bDDt}[1]{\frac{\rD #1}{\rD t}}

\newcommand{\pd}[2]{\frac{\partial#1}{\partial#2}}

\title{Dynamics of tsunami waves}
\author{\href{http://www.cmla.ens-cachan.fr/\~dias}{Frédéric Dias%
\footnote{Centre de Mathématiques et de Leurs Applications, 
\'{E}cole Normale Supérieure de Cachan,
61, avenue du Président Wilson, 
94235 Cachan cedex, France}} \and
\href{http://www.cmla.ens-cachan.fr/\~dutykh}{Denys Dutykh\footnotemark[1]}}
\date{}

\begin{document}

\maketitle

\begin{abstract}
The life of a tsunami is usually divided into three phases: the generation
(tsunami source), the propagation and the inundation. Each phase is complex
and often described separately. A brief description of each phase is given.
Model problems are identified. Their formulation is given. While some of these 
problems can be solved analytically, most require numerical techniques. The inundation phase 
is less documented than the other phases. It is shown that methods based on Smoothed
Particle Hydrodynamics (SPH) are particularly well-suited for the inundation phase. 
Directions for future research are outlined.
\end{abstract}

\tableofcontents

\section{Introduction}

Given the broadness of the topic of tsunamis, our purpose here is to recall some of the
basics of tsunami modeling and to emphasize some
general aspects, which are sometimes overlooked. The life of a tsunami is usually divided into 
three phases: the generation (tsunami source), the propagation and the inundation. The third 
and most difficult phase of the dynamics of tsunami waves deals with their breaking as they approach 
the shore. This phase depends greatly on the bottom bathymetry and on the coastline type.
The breaking can be progressive. Then the inundation process is relatively slow and can last for
several minutes. Structural damages are mainly caused by inundation. The breaking can also be
explosive and lead to the formation of a plunging jet. The impact on the coast is then very
rapid. In very shallow water, the amplitude of tsunami waves
grows to such an extent that typically an undulation appears on the long wave, which develops into 
a progressive bore \cite{Chanson05}.  This turbulent front, similar to the wave that occurs when
a dam breaks, can be quite high and travel onto the beach at great speed.  Then the front and the turbulent 
current behind it move onto the shore, past buildings 
and vegetation until they are finally stopped by rising ground. The water level can rise rapidly,
typically from 0 to 3 meters in 90 seconds. 

The trajectory of these currents and their velocity are quite unpredictible, especially in the final 
stages because they are sensitive to small changes in the topography, and to 
the stochastic patterns of the collapse of buildings, and to the accumulation of debris such as 
trees, cars, logs, furniture.
The dynamics of this final stage of tsunami waves is somewhat similar to the dynamics of flood waves 
caused by dam breaking, dyke breaking or overtopping of dykes (cf. the recent tragedy of
hurricane Katrina in August 2005). Hence 
research on flooding events and measures to deal with them may be able to contribute to improved 
warning and damage reduction systems for tsunami waves in the areas of the world where these waves 
are likely to occur as shallow surge waves (cf. the recent tragedy of the Indian Ocean
tsunami in December 2004).

Civil engineers who visited the damage area following the Boxing day tsunami came up with several basic 
conclusions. Buildings that had been constructed to satisfy modern safety standards offered a satisfactory 
resistance, in particular those with reinforced concrete beams properly integrated in the frame structure.  
These were able to withstand pressure associated with the leading front of the order of 1 
atmosphere (recall that an equivalent pressure $p$ is obtained with a windspeed $U$ of about 450 m/s, since
$p=\rho_{\rm air}U^2/2$).  
By contrast brick buildings collapsed and were washed away.  Highly porous or open structures survived. 
Buildings further away from the beach survived the front in some cases, but 
they were then destroyed by the erosion of the ground around the buildings by the water currents \cite{Hunt05}.

Section 2 provides a description of the tsunami source when the source is an earthquake. In Section 3,
we review the equations that are often used for tsunami propagation. Section 4 provides a short discussion
on the energy of tsunamis. Section 5 is devoted to the run-up and inundation of tsunamis. Finally
directions for future research are outlined.

\section{Tsunami induced by near-shore earthquake}

The inversion of seismic data allows one to reconstruct the permanent deformations of the sea
bottom following earthquakes. In spite of the complexity of the seismic source and of the internal
structure of the Earth, scientists have been relatively successful in using simple models for the 
source \cite{Okada85}. A description of Okada's model follows. 

\subsection{Introduction}

The fracture zones, along which the foci of earthquakes are to be
found, have been described in various papers. For example, it has been
suggested that Volterra's theory of dislocations might be the proper tool for a quantitative
description of these fracture zones \cite{stek2}. This suggestion was made
for the following reason. If the mechanism involved in earthquakes and the
fracture zones is indeed one of fracture, 
discontinuities in the displacement components across the fractured
surface will exist. As dislocation theory may be described as that part of the
theory of elasticity dealing with surfaces across which the
displacement field is discontinuous, the suggestion seems
reasonable.

As commonly done in mathematical physics, it is necessary for simplicity's
sake to make some assumptions. Here we neglect
the curvature of the earth, its gravity, temperature, magnetism,
non-homogeneity, and consider a semi-infinite medium, which is
homogeneous and isotropic. We further assume that the laws of
classical linear elasticity theory hold. 

Several studies showed that the effect of earth curvature is negligible 
for shallow events at distances of less than $20^\circ$ \cite{mcginl,ben1,ben2,smylie}. 
The sensitivity to earth topography, homogeneity, isotropy and half-space assumptions
was studied and discussed recently \cite{tim}. The author
used a commercially available code, ABACUS, which is based on a finite element model (FEM). 
Six FEMs were constructed to test the sensitivity of
deformation predictions to each assumption. The main conclusion is
that the vertical layering of lateral inhomogeneity can sometimes
cause considerable effects on the deformation fields.

The usual boundary conditions for dealing with earth's problems
require that the surface $S$ of the elastic medium (the earth) shall
be free from forces. The resulting mixed boundary-value problem was solved a century ago
\cite{volt}. Later, Steketee proposed an alternative
method to solve this problem using Green's functions \cite{stek2}.

\subsection{Volterra's theory of dislocations}

In order to introduce the concept of dislocation and for simplicity's sake,
this section is devoted to the case of an entire elastic space. The
second reason is that in his original paper Volterra solved
the problem in this case \cite{volt}.

Let $O$ be the origin of a Cartesian coordinate system in an
infinite elastic medium, $x_i$ the Cartesian coordinates
$(i=1,2,3)$, and $\mathbf{e}_i$ a unit vector in the positive
$x_i$-direction. A force $\mathbf{F}=F \mathbf{e}_k$ at $O$
generates a displacement field $u_i^k(P,O)$ at point $P$, which is determined by
the well-known Somigliana tensor
\begin{equation}\label{somigliana}
  u_i^k(P,O) = \frac{F}{8\pi\mu} (\delta_{ik} r_{,\: nn}
  - \alpha r_{,\:ik}), \quad \mbox{with} \;\; \alpha=\frac{\lambda+\mu}{\lambda+2\mu}.
\end{equation}
In this relation $\delta_{ik}$ is the Kronecker delta,
$\lambda$ and $\mu$ are Lam\'e's constants, and
$r$ is the distance from $P$ to $O$. The coefficient $\alpha$ can be rewritten as
$\alpha=1/2(1-\nu)$, where $\nu$ is Poisson's ratio. Later we will also use Young's modulus
$E$, which is defined as
\[ E = \frac{\mu\,(3\lambda+2\mu)}{\lambda+\mu}. \]
The notation 
$r_{,\:i}$ means $\partial r/\partial x_i$ and the summation convention
applies.

The stresses due to the displacement field (\ref{somigliana}) are
easily computed from Hooke's law:

\begin{equation}\label{hook}
  \sigma_{ij} = \lambda\delta_{ij} u_{k,k} + \mu
  (u_{i,j}+u_{j,i}).
\end{equation}
We find
\[
  \sigma_{ij}^k (P,O) = -\frac{\alpha F}{4\pi} \left(
  3\frac{x_i x_j x_k}{r^5} + \frac{\mu}{\lambda+\mu}
  \frac{\delta_{ki}x_j + \delta_{kj}x_i - \delta_{ij}x_k}{r^3}
  \right).
\]
The components of the force per unit area on a
surface element are denoted as follows: 
\[
  T_i^{ k} = \sigma_{ij}^k\cdot\nu_j,
\]
where the $\nu_j$'s are the direction cosines of the normal to the
surface element \cite{sokol}. A Volterra dislocation is defined as a surface
$\Sigma$ in the elastic medium across which there is a discontinuity
$\Delta u_i$ in the displacement fields of the type
\begin{eqnarray}\label{displdef}
  \Delta u_i & = & u_i^+ - u_i^- = U_i + \Omega_{ij}x_j, \\
  \Omega_{ij}& = & -\Omega_{ji}.
\end{eqnarray}
Equation (\ref{displdef}) in which $U_i$ and $\Omega_{ij}$ are
constants is the well-known Weingarten relation which states that
the discontinuity $\Delta u_i$ should be of the type of a rigid body
displacement, thereby maintaining continuity of the components of
stress and strain across $\Sigma$.

The displacement field in an infinite elastic medium due to a
dislocation of type (\ref{somigliana}) is then determined by
Volterra's formula \cite{volt}
\begin{equation}\label{volt2}
  u_k(y_1,y_2,y_3) := u_k(y_l) = \frac{1}{F}\int\!\!\!\!\int
  \limits_{\!\!\!\!\!\!\!\Sigma} \Delta u_i
  T_i^{ k} dS.
\end{equation}

Once the surface $\Sigma$ is given, the dislocation is essentially
determined by the six constants $U_i$ and $\Omega_{ij}$. Therefore we also write
\begin{equation}\label{eldisloc}
  u_k(y_l) = \frac{U_i}{F}\int\!\!\!\!\int
  \limits_{\!\!\!\!\!\!\!\Sigma} \sigma_{ij}^k (P,Q) \nu_j dS +
  \frac{\Omega_{ij}}{F}\int\!\!\!\!\int
  \limits_{\!\!\!\!\!\!\!\Sigma} \{x_j\sigma_{il}^k(P,Q) - x_i
  \sigma_{jl}^k(P,Q)\}\nu_l dS,
\end{equation}
where $\Omega_{ij}$ takes only the values $\Omega_{12}$,
$\Omega_{23}$, $\Omega_{31}$. Following Volterra \cite{volt} and
Love \cite{love} we call each of the six integrals in
(\ref{eldisloc}) an elementary dislocation.

It is clear from (\ref{volt2}) and (\ref{eldisloc}) that the
computation of the displacement field $u_k(Q)$ is performed as
follows: A force $F\mathbf{e}_k$ is applied at $Q$, and the stresses
$\sigma_{ij}^k(P,Q)$ that this force generates are computed at the
points $P(x_i)$ on $\Sigma$. In particular the components of the
force on $\Sigma$ are computed. After multiplication with prescribed
weights of magnitude $\Delta u_i$ these forces are integrated over
$\Sigma$ to give the displacement component in $Q$ due to the
dislocation on $\Sigma$.

\subsection{Dislocations in elastic half-space}

When the case of an elastic half-space is considered, equation
(\ref{volt2}) remains valid, but we have to replace $\sigma_{ij}^k$ by
another tensor $\omega_{ij}^k$. This can be explained by the fact
that the elementary solutions for a half-space are different from
Somigliana solution (\ref{somigliana}).

The $\omega_{ij}^k$ can be obtained from the displacements
corresponding to nuclei of strain in a half-space through relation
(\ref{hook}). Steketee showed a method of obtaining the six
$\omega_{ij}^k$ fields by using a Green's function and derived
$\omega_{12}^k$, which is relevant to a vertical strike-slip
fault. Maruyama derived the remaining five functions \cite{maru}.

It is interesting to mention here that historically these solutions were first
derived in a straightforward manner by Mindlin \cite{mindl1,mindl2},
who gave explicit expressions of the displacement and stress
fields for half-space nuclei of strain consisting of single forces
with and without moment. It is only necessary to write the single
force results since the other forms can be obtained by taking
appropriate derivatives. Their method consists in finding the
displacement field in Westergaard's form of the Galerkin vector \cite{wester}. 
This vector is then determined by taking a linear
combination of some biharmonic elementary solutions. The
coefficients are chosen to satisfy boundary and equilibrium
conditions. These solutions were also derived by Press 
in a slightly different manner \cite{press}.

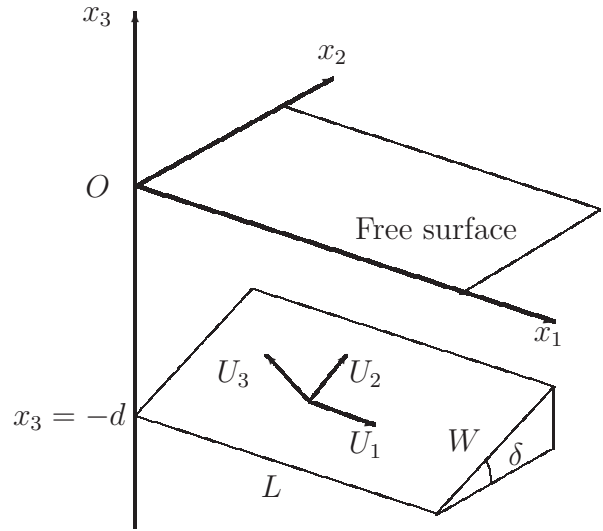
\begin{figure}[htbp]
\begin{center}
\unitlength 1mm
\begin{picture}(71.25,68.75)(0,0)

\linethickness{0.50mm}
\put(9.38,0.00){\line(0,1){68.75}}
\put(9.38,68.75){\vector(0,1){0.12}}

\put(4.38,68.13){\makebox(0,0)[cc]{$x_3$}}

\linethickness{0.50mm}
\multiput(9.38,45.63)(0.37,-0.12){151}{\line(1,0){0.37}}
\put(65.00,27.50){\vector(3,-1){0.12}}

\linethickness{0.15mm}
\multiput(9.38,15.00)(0.37,-0.12){109}{\line(1,0){0.37}}

\put(64.38,25.63){\makebox(0,0)[cc]{$x_1$}}

\linethickness{0.50mm}
\multiput(9.38,45.63)(0.22,0.12){120}{\line(1,0){0.22}}
\put(35.63,60.00){\vector(2,1){0.12}}

\put(35.63,63.13){\makebox(0,0)[cc]{$x_2$}}

\put(4.38,45.63){\makebox(0,0)[cc]{$O$}}

\put(0.63,15.00){\makebox(0,0)[cc]{$x_3=-d$}}

\linethickness{0.15mm}
\multiput(49.38,1.87)(0.21,0.12){73}{\line(1,0){0.21}}

\linethickness{0.15mm}
\put(65.00,10.63){\line(0,1){8.13}}

\linethickness{0.15mm}
\multiput(49.38,1.88)(0.12,0.13){130}{\line(0,1){0.13}}

\linethickness{0.15mm}
\multiput(25.00,31.88)(0.37,-0.12){109}{\line(1,0){0.37}}

\linethickness{0.15mm}
\multiput(9.38,15.00)(0.12,0.13){130}{\line(0,1){0.13}}

\put(60.00,10.00){\makebox(0,0)[cc]{$\delta$}}

\linethickness{0.15mm}
\multiput(56.74,7.12)(0.12,-0.99){1}{\line(0,-1){0.99}}
\multiput(56.50,8.07)(0.12,-0.48){2}{\line(0,-1){0.48}}
\multiput(56.15,8.96)(0.12,-0.30){3}{\line(0,-1){0.30}}

\put(27.50,5.63){\makebox(0,0)[cc]{$L$}}

\put(53.13,11.88){\makebox(0,0)[cc]{$W$}}

\linethickness{0.45mm}
\multiput(32.50,16.88)(0.34,-0.12){26}{\line(1,0){0.34}}
\put(41.25,13.75){\vector(3,-1){0.12}}

\linethickness{0.45mm}
\multiput(32.50,16.88)(0.12,0.15){42}{\line(0,1){0.15}}
\put(37.50,23.13){\vector(3,4){0.12}}

\put(40.00,11.25){\makebox(0,0)[cc]{$U_1$}}

\put(40.00,20.63){\makebox(0,0)[cc]{$U_2$}}

\linethickness{0.45mm}
\multiput(26.88,23.13)(0.12,-0.13){47}{\line(0,-1){0.13}}
\put(26.88,23.13){\vector(-1,1){0.12}}

\put(22.50,20.63){\makebox(0,0)[cc]{$U_3$}}

\linethickness{0.20mm}
\multiput(28.75,56.25)(0.37,-0.12){115}{\line(1,0){0.37}}

\linethickness{0.20mm}
\multiput(52.50,31.25)(0.20,0.12){94}{\line(1,0){0.20}}

\put(49.38,40.00){\makebox(0,0)[cc]{Free surface}}

\end{picture}
\end{center}
  \caption{Coordinate system adopted in this study and
geometry of the source model}\label{fig:okad}
\end{figure}

Here, we take the Cartesian coordinate system shown in
Figure~\ref{fig:okad}. The elastic medium occupies the region $z\leq
0$ and the $x-$axis is taken to be parallel to the strike direction of
the fault. In this coordinate system,
$u_i^j(x_1,x_2,x_3;\xi_1,\xi_2,\xi_3)$ is the $i$th component of the
displacement at $(x_1,x_2,x_3)$ due to the $j$th direction point
force of magnitude $F$ at $(\xi_1,\xi_2,\xi_3)$. It can be expressed as
follows \cite{mindl1,press,Okada85,okada92}:
\begin{equation}\label{okada1}
  u_i^j (x_1,x_2,x_3) = u_{iA}^j(x_1,x_2,-x_3) -
  u_{iA}^j(x_1,x_2,x_3) + u_{iB}^j(x_1,x_2,x_3) + x_3
  u_{iC}^j(x_1,x_2,x_3),
\end{equation}
where
\begin{eqnarray*}
  u_{iA}^j & = & \frac{F}{8\pi\mu}\left((2-\alpha)\frac{\delta_{ij}}{R}
  + \alpha \frac{R_iR_j}{R^3}\right), \\
  u_{iB}^j & = & \frac{F}{4\pi\mu}\Biggl(
  \frac{\delta_{ij}}{R} + \frac{R_iR_j}{R^3} +
  \frac{1-\alpha}{\alpha}\Bigl[
  \frac{\delta_{ij}}{R+R_3} +\\ & & + \frac{R_i\delta_{j3}-
  R_j\delta_{i3}(1-\delta_{j3})}{R(R+R_3)} -
  \frac{R_iR_j}{R(R+R_3)^2}(1-\delta_{i3})(1-\delta_{j3})
  \Bigr]\Biggr), \\
    u_{iC}^j & = & \frac{F}{4\pi\mu}(1-2\delta_{i3})
    \left(
    (2-\alpha)\frac{R_i\delta_{j3}-R_j\delta_{i3}}{R^3} +
    \alpha\xi_3\left[
    \frac{\delta_{ij}}{R^3} - 3\frac{R_iR_j}{R^5}
    \right]\right).
\end{eqnarray*}
In these expressions $R_1 = x_1-\xi_1$, $R_2=x_2-\xi_2$, $R_3=x_3-\xi_3$, $R^2 =
R_1^2 + R_2^2 + R_3^2$.

The first term in equation (\ref{okada1}), $u_{iA}^j(x_1,x_2,x_3)$, is the well-known 
Somigliana tensor, which
represents the displacement field due to a single force placed at
$(\xi_1,\xi_2,\xi_3)$ in an infinite medium \cite{love}. The second
term also looks like a Somigliana tensor. This term corresponds to a
contribution from an image source of the given point force placed at
$(\xi_1,\xi_2,-\xi_3)$ in the infinite medium. The third term,
$u_{iB}^j(x_1,x_2,x_3)$, and $u_{iC}^j(x_1,x_2,x_3)$ in the fourth
term are naturally depth dependent. When $x_3$ is set equal to zero in
equation (\ref{okada1}), the first and the second terms cancel each
other, and the fourth term vanishes. The remaining term,
$u_{iB}^j(x_1,x_2,0)$ reduces to the formula for the surface
displacement field due to a point force in a half-space
\cite{Okada85}:
\[
\left\{%
\begin{array}{ll}
    u_1^1 = \frac{F}{4\pi\mu}\left(
    \frac1R + \frac{(x_1-\xi_1)^2}{R^3} + \frac{\mu}{\lambda+\mu}
    \left[
    \frac{1}{R-\xi_3} - \frac{(x_1-\xi_1)^2}{R(R-\xi_3)^2}\right]\right), & \\
    u_2^1 = \frac{F}{4\pi\mu}(x_1-\xi_1)(x_2-\xi_2)\left(
    \frac1{R^3} - \frac{\mu}{\lambda+\mu}\frac1{R(R-\xi_3)^2}
    \right), &  \\
    u_3^1 = \frac{F}{4\pi\mu}(x_1-\xi_1)\left(
    -\frac{\xi_3}{R^3} - \frac{\mu}{\lambda+\mu} \frac1{R(R-\xi_3)}
    \right), &
\end{array}%
\right.
\]

\[
\left\{%
\begin{array}{ll}
    u_1^2 = \frac{F}{4\pi\mu}(x_1-\xi_1)(x_2-\xi_2)\left(
    \frac1{R^3} - \frac{\mu}{\lambda+\mu}\frac{1}{R(R-\xi_3)^2}\right), & \\
    u_2^2 = \frac{F}{4\pi\mu}\left(
    \frac1R + \frac{(x_2-\xi_2)^2}{R^3} + \frac{\mu}{\lambda+\mu}
    \left[
    \frac{1}{R-\xi_3} - \frac{(x_2-\xi_2)^2}{R(R-\xi_3)^2}\right]\right), &  \\
    u_3^2 = \frac{F}{4\pi\mu}(x_2-\xi_2)\left(
    -\frac{\xi_3}{R^3} - \frac{\mu}{\lambda+\mu} \frac1{R(R-\xi_3)}
    \right), &  
\end{array}%
\right.
\]

\[
\left\{%
\begin{array}{ll}
    u_1^3 = \frac{F}{4\pi\mu}(x_1-\xi_1)\left(
    -\frac{\xi_3}{R^3} + \frac{\mu}{\lambda+\mu}\frac{1}{R(R-\xi_3)}\right), & \\
    u_2^3 = \frac{F}{4\pi\mu}(x_2-\xi_2)\left(
    -\frac{\xi_3}{R^3} + \frac{\mu}{\lambda+\mu}\frac{1}{R(R-\xi_3)}\right), &  \\
    u_3^3 = \frac{F}{4\pi\mu}\left(
    \frac1{R} + \frac{\xi_3^2}{R^3} + \frac{\mu}{\lambda+\mu}\frac1R
    \right). &  
\end{array}%
\right.
\]
In these formulas $R^2 = (x_1-\xi_1)^2 + (x_2-\xi_2)^2 + \xi_3^2$.

In order to obtain the displacements due to the
dislocation we need to calculate the corresponding $\xi_k$-derivatives
of the point force solution (\ref{okada1}) and to put it in
Steketee-Volterra formula (\ref{volt2}) 
\[
  u_i = \frac1F\int\!\!\!\!\int
  \limits_{\!\!\!\!\!\!\!\Sigma} \Delta u_j
  \left[
  \lambda\delta_{jk} \pd{u_i^n}{\xi_n} +
  \mu\left(\pd{u_i^j}{\xi_k} + \pd{u_i^k}{\xi_j}
  \right)\right]\nu_k dS.
\]
It is expressed as follows:
\begin{eqnarray*}
  \pd{u_i^j}{\xi_k} (x_1,x_2,x_3) & = & \pd{u_{iA}^j}{\xi_k}(x_1,x_2,-x_3) -
  \pd{u_{iA}^j}{\xi_k}(x_1,x_2,x_3) +\\ & & + \pd{u_{iB}^j}{\xi_k}(x_1,x_2,x_3) + x_3
  \pd{u_{iC}^j}{\xi_k}(x_1,x_2,x_3),
\end{eqnarray*}
with
\begin{eqnarray*}
  \pd{u_{iA}^j}{\xi_k} & = & \frac{F}{8\pi\mu}\left(
  (2-\alpha)\frac{R_k}{R^3}\delta_{ij} - \alpha
  \frac{R_i\delta_{jk}+R_j\delta_{ik}}{R^3} + 3\alpha
  \frac{R_i R_j R_k}{R^5}
  \right), \\
  \pd{u_{iB}^j}{\xi_k} & = & \frac{F}{4\pi\mu}\left(
  -\frac{R_i\delta_{jk} + R_j\delta_{ik} - R_k\delta_{ij}}{R^3} +
  3\frac{R_iR_jR_k}{R^5} \right. +\\ & & + \frac{1-\alpha}{\alpha}\Bigl[
  \frac{\delta_{3k}R + R_k}{R(R+R_3)^2}\delta_{ij} -
  \frac{\delta_{ik}\delta_{j3} -
  \delta_{jk}\delta_{i3}(1-\delta_{j3})}{R(R+R_3)} +\\ & & +
  \bigl(R_i\delta_{j3} - R_j\delta_{i3}(1-\delta_{j3})\bigr)
  \frac{\delta_{3k}R^2+R_k(2R+R_3)}{R^3(R+R_3)^2} +\\ & & \left. +
  (1-\delta_{i3})(1-\delta_{j3})\bigl(
  \frac{R_i\delta_{jk}+R_j\delta_{ik}}{R(R+R_3)^2} -
  R_iR_j\frac{2\delta_{3k}R^2 + R_k(3R+R_3)}{R^3(R+R_3)^3}
  \bigr)\Bigr]\right), \\
  \pd{u_{iC}^j}{\xi_k} & = & \frac{F}{4\pi\mu}(1-2\delta_{i3})\biggl(
  (2-\alpha)\Bigl[
  \frac{\delta_{jk}\delta_{i3}-\delta_{ik}\delta_{j3}}{R^3} +
  \frac{3R_k(R_i\delta_{j3}-R_j\delta_{i3})}{R^5}\Bigr] + \\& & +
  \alpha\delta_{3k}\Bigl[\frac{\delta_{ij}}{R^3} -
  \frac{3R_iR_j}{R^5}\Bigr] + 3\alpha\xi_3\Bigl[
  \frac{R_i\delta_{jk}+R_j\delta_{ik}+R_k\delta_{ij}}{R^5} -
  \frac{5R_iR_jR_k}{R^7}\Bigr]\biggr).
\end{eqnarray*}

\subsection{Finite rectangular source}
Now, let us consider a more practical problem. We define the elementary
dislocations $U_1$, $U_2$, and $U_3$, corresponding to the
strike-slip, dip-slip, and tensile components of an arbitrary
dislocation. In Figure \ref{fig:okad} each vector represents the
direction of the elementary faults. The vector $\Dv$ is the 
so-called Burger's vector, which shows how both sides of the fault are spread
out: $\Dv = \uv^+ - \uv^-$.

\begin{figure}[htbp]
\begin{center}
\unitlength 1mm
\def\Dv{\mathbf{D}}

\begin{picture}(83.75,84.38)(0,0)

\linethickness{0.55mm}
\put(10.00,54.38){\line(1,0){70.00}}
\put(80.00,54.38){\vector(1,0){0.12}}

\linethickness{0.55mm}
\multiput(10.00,54.38)(0.16,0.12){167}{\line(1,0){0.16}}
\put(36.25,74.38){\vector(4,3){0.12}}

\put(83.75,54.38){\makebox(0,0)[cc]{$x$}}

\put(33.13,77.50){\makebox(0,0)[cc]{$y$}}

\linethickness{0.45mm}
\put(10.00,-0.63){\line(0,1){85.00}}
\put(10.00,84.38){\vector(0,1){0.12}}

\put(6.25,83.75){\makebox(0,0)[cc]{$z$}}

\put(5.63,54.38){\makebox(0,0)[cc]{$O$}}

\linethickness{0.15mm}
\multiput(10.00,10.00)(0.12,0.31){89}{\line(0,1){0.31}}

\linethickness{0.15mm}
\put(10.00,10.00){\line(1,0){54.38}}

\linethickness{0.15mm}
\multiput(64.38,10.00)(0.12,0.31){89}{\line(0,1){0.31}}

\linethickness{0.15mm}
\multiput(64.38,10.00)(0.16,0.12){68}{\line(1,0){0.16}}

\linethickness{0.15mm}
\put(75.01,18.13){\line(0,1){19.37}}

\put(68.76,15.63){\makebox(0,0)[cc]{$\delta$}}

\linethickness{0.15mm}
\multiput(66.37,12.73)(0.12,-0.12){6}{\line(1,0){0.12}}
\multiput(65.42,13.10)(0.32,-0.12){3}{\line(1,0){0.32}}

\linethickness{0.25mm}
\multiput(30.63,18.75)(0.12,0.16){83}{\line(0,1){0.16}}
\put(40.63,31.88){\vector(3,4){0.12}}

\linethickness{0.15mm}
\multiput(30.63,18.75)(1.84,0.79){10}{\multiput(0,0)(0.31,0.13){3}{\line(1,0){0.31}}}

\put(25.01,13.13){\makebox(0,0)[cc]{Fault plane}}

\put(31.88,58.76){\makebox(0,0)[cc]{Free surface}}

\put(42.51,30.63){\makebox(0,0)[cc]{$\Dv$}}

\linethickness{0.15mm}
\put(21.26,18.75){\line(1,0){36.87}}
\put(58.13,18.75){\vector(1,0){0.12}}

\put(55.01,16.25){\makebox(0,0)[cc]{$x'$}}

\linethickness{0.15mm}
\multiput(35.93,21.84)(0.13,-0.25){2}{\line(0,-1){0.25}}
\multiput(35.47,22.29)(0.12,-0.11){4}{\line(1,0){0.12}}
\multiput(34.84,22.66)(0.21,-0.12){3}{\line(1,0){0.21}}
\multiput(34.08,22.93)(0.38,-0.14){2}{\line(1,0){0.38}}

\put(37.51,23.13){\makebox(0,0)[cc]{$\phi$}}

\linethickness{0.15mm}
\multiput(35.55,19.64)(0.08,-0.68){1}{\line(0,-1){0.68}}
\multiput(35.26,20.29)(0.15,-0.32){2}{\line(0,-1){0.32}}
\multiput(34.76,20.86)(0.12,-0.14){4}{\line(0,-1){0.14}}

\linethickness{0.15mm}
\multiput(36.00,19.81)(0.12,-0.57){2}{\line(0,-1){0.57}}
\multiput(35.53,20.78)(0.12,-0.24){4}{\line(0,-1){0.24}}

\put(39.38,20.00){\makebox(0,0)[cc]{$\theta$}}

\put(49.38,14.38){\makebox(0,0)[cc]{}}

\linethickness{0.15mm}
\multiput(30.63,18.75)(0,1.92){8}{\line(0,1){0.96}}

\put(31.88,7.50){\makebox(0,0)[cc]{$L$}}

\put(66.26,26.25){\makebox(0,0)[cc]{$W$}}

\linethickness{0.15mm}
\put(20.63,37.50){\line(1,0){54.38}}

\end{picture}
\end{center}
  \caption{Geometry of the source model and orientation
  of Burger's vector $\Dv$}\label{fig:2}
\end{figure}
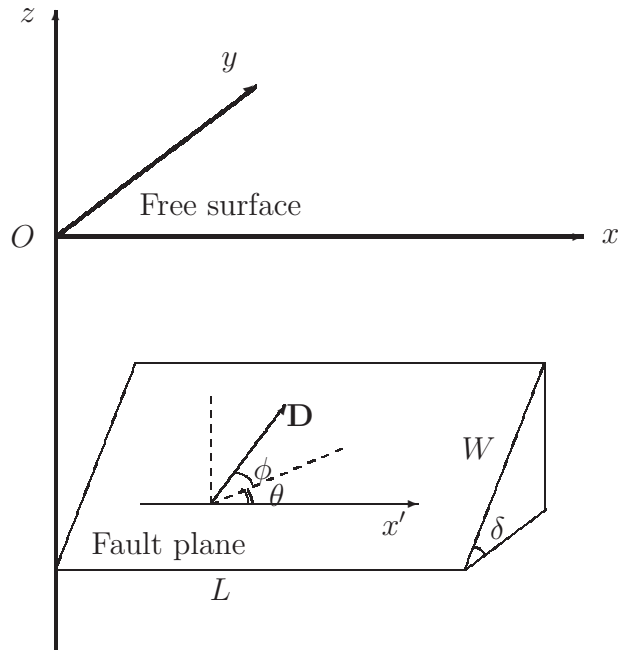

A general dislocation can be determined by three angles:
the dip angle $\delta$ of the fault, the slip angle $\theta$, and
the angle $\phi$ between the fault plane and Burger's vector
$\Dv$. This situation is schematically described in
Figure \ref{fig:2}.

For a finite rectangular fault with length $L$ and width $W$ occurring at depth $d$ (Figure
\ref{fig:2}), the deformation field can be evaluated analytically by
changing the variables and performing integration over the
rectangle. This was done by several authors
\cite{chin,sat74,iwa79,Okada85,okada92}. Here we give the results of
their computations. The final results represented in compact form
are listed below using Chinnery's notation $\|$ to
represent the substitution
\[
  f(\xi,\eta)\| = f(x,p) - f(x,p-W) - f(x-L,p) + f(x-L,p-W).
\]

Let us introduce the following notation:
\[
  p = y\cos\delta + d\sin\delta, \quad
  q = y\sin\delta - d\cos\delta,
\]
\[
  \tilde{y} = \eta\cos\delta + q\sin\delta, \quad
  \tilde{d} = \eta\sin\delta - q\cos\delta,
\]
\[
  R^2 = \xi^2 + \eta^2 + q^2 = \xi^2 + \tilde{y}^2 + \tilde{d}^2, \quad
  X^2 = \xi^2 + q^2.
\]

The quantities $U_1$, $U_2$ and $U_3$ are linked to Burger's
vector through the identities
\[
  U_1 = |\Dv| \cos\phi\cos\theta, \quad
  U_2 = |\Dv| \cos\phi\sin\theta, \quad
  U_3 = |\Dv| \sin\phi.
\]
For a strike-slip dislocation, one has
\begin{eqnarray*}
  u_1 & = & -\frac{U_1}{2\pi}\left.\left(
  \frac{\xi q}{R(R+\eta)} + \arctan\frac{\xi\eta}{qR} +
  I_1\sin\delta\right)\right\|, \\
  u_2 & = & -\frac{U_1}{2\pi}\left.\left(
  \frac{\tilde{y} q}{R(R+\eta)} + \frac{q\cos\delta}{R+\eta} +
  I_2\sin\delta\right)\right\|, \\
  u_3 & = & -\frac{U_1}{2\pi}\left.\left(
  \frac{\tilde{d} q}{R(R+\eta)} + \frac{q\sin\delta}{R+\eta} +
  I_4\sin\delta\right)\right\|.
\end{eqnarray*}
For a dip-slip dislocation, one has
\begin{eqnarray*}
  u_1 & = & -\frac{U_2}{2\pi}\left.\left(
  \frac{q}{R} - I_3\sin\delta\cos\delta
  \right)\right\|, \\
  u_2 & = & -\frac{U_2}{2\pi}\left.\left(
  \frac{\tilde{y} q}{R(R+\xi)} +
  \cos\delta\arctan\frac{\xi\eta}{qR} - I_1\sin\delta\cos\delta
  \right)\right\|, \\
  u_3 & = & -\frac{U_2}{2\pi}\left.\left(
  \frac{\tilde{d} q}{R(R+\xi)} +
  \sin\delta\arctan\frac{\xi\eta}{qR} - I_5\sin\delta\cos\delta
  \right)\right\|.
\end{eqnarray*}
For a tensile fault dislocation, one has
\begin{eqnarray*}
  u_1 & = & \frac{U_3}{2\pi}\left.\left(
  \frac{q^2}{R(R+\eta)} - I_3\sin^2\delta
  \right)\right\|, \\
  u_2 & = & \frac{U_3}{2\pi}\left.\left(
  \frac{-\tilde{d} q}{R(R+\xi)} - \sin\delta\left[
  \frac{\xi q}{R(R+\eta)} - \arctan\frac{\xi\eta}{qR}
  \right] - I_1\sin^2\delta
  \right)\right\|, \\
  u_3 & = & \frac{U_3}{2\pi}\left.\left(
  \frac{\tilde{y} q}{R(R+\xi)} + \cos\delta\left[
  \frac{\xi q}{R(R+\eta)} - \arctan\frac{\xi\eta}{qR}
  \right] - I_5\sin^2\delta
  \right)\right\|.
\end{eqnarray*}
The terms $I_1,\dots,I_5$ are given by 
\begin{eqnarray*}
  I_1 & = &
  -\frac{\mu}{\lambda+\mu}\frac{\xi}{(R+\tilde{d})\cos\delta} -
  \tan\delta I_5, \\
  I_2 & = & -\frac{\mu}{\lambda+\mu}\log(R+\eta) - I_3, \\
  I_3 & = & \frac{\mu}{\lambda+\mu}\left[
  \frac1{\cos\delta}\frac{\tilde{y}}{R+\tilde{d}} - \log(R+\eta)
  \right] + \tan\delta I_4, \\
  I_4 & = & \frac{\mu}{\mu+\lambda}\frac1{\cos\delta}\left(
  \log(R+\tilde{d}) - \sin\delta \log(R+\eta)
  \right), \\
  I_5 & = & \frac{\mu}{\lambda+\mu}\frac2{\cos\delta}
  \arctan\frac{\eta(X+q\cos\delta)+X(R+X)\sin\delta}{\xi(R+X)\cos\delta},
\end{eqnarray*}
and if $\cos\delta=0$,
\begin{eqnarray*}
  I_1 & = & -\frac{\mu}{2(\lambda+\mu)}
  \frac{\xi q}{(R+\tilde{d})^2}, \\
  I_3 & = & \frac{\mu}{2(\lambda+\mu)} \left[
  \frac{\eta}{R+\tilde{d}} + \frac{\tilde{y} q}{(R+\tilde{d})^2} -
  \log(R+\eta)\right], \\
  I_4 & = & -\frac{\mu}{\lambda+\mu} \frac{q}{R+\tilde{d}}, \\
  I_5 & = & -\frac{\mu}{\lambda+\mu} \frac{\xi\sin\delta}{R+\tilde{d}}.
\end{eqnarray*}

\begin{figure}
  \includegraphics[width=\linewidth]{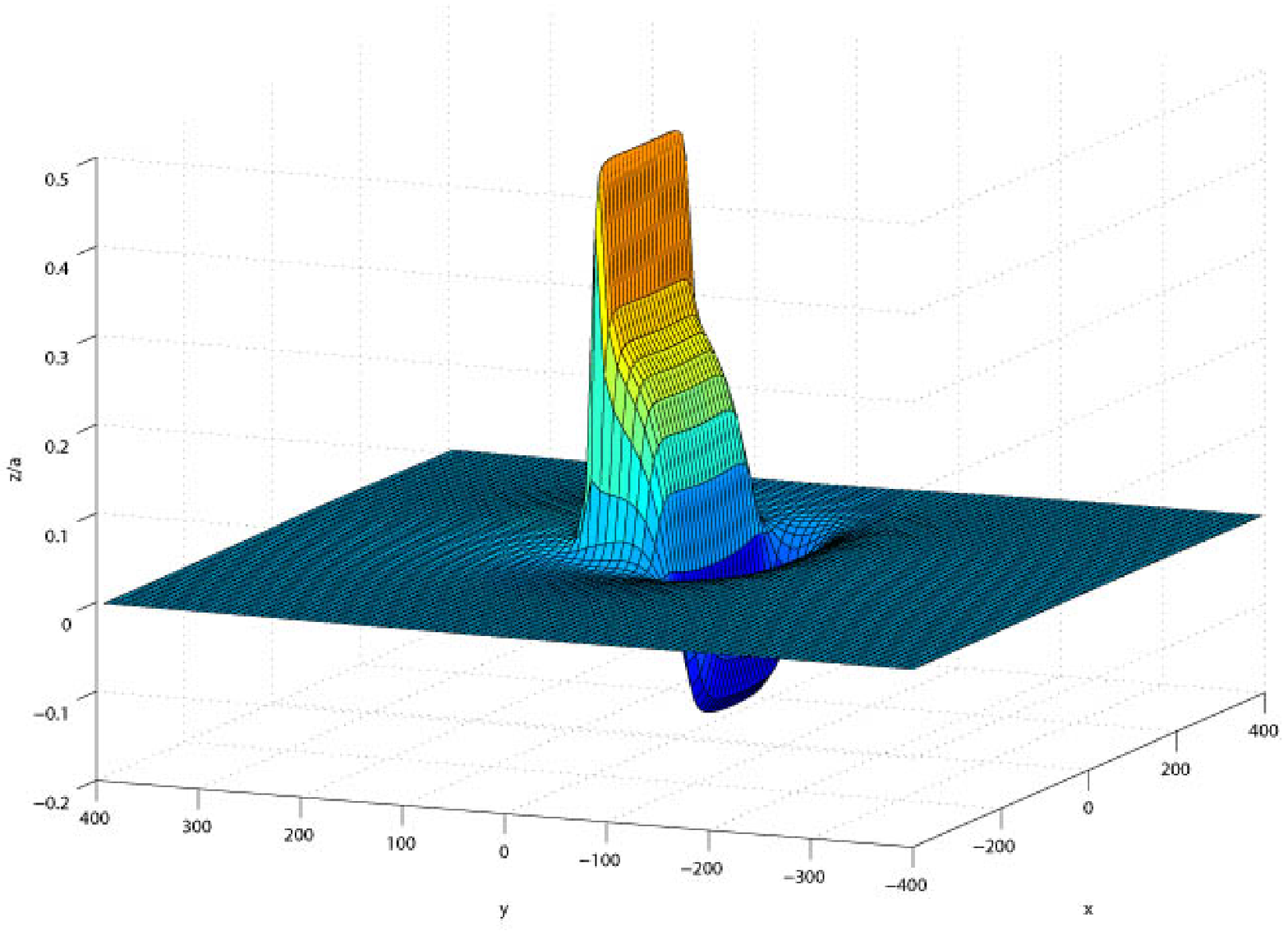}\\
  \caption{Dimensionless free-surface deformation $z/a$ after dip-slip fault. Here $a$
is $|\Dv|$ (30 m in the present application).}
  \label{fig:dip}
\end{figure}

\begin{figure}
  \includegraphics[width=\linewidth]{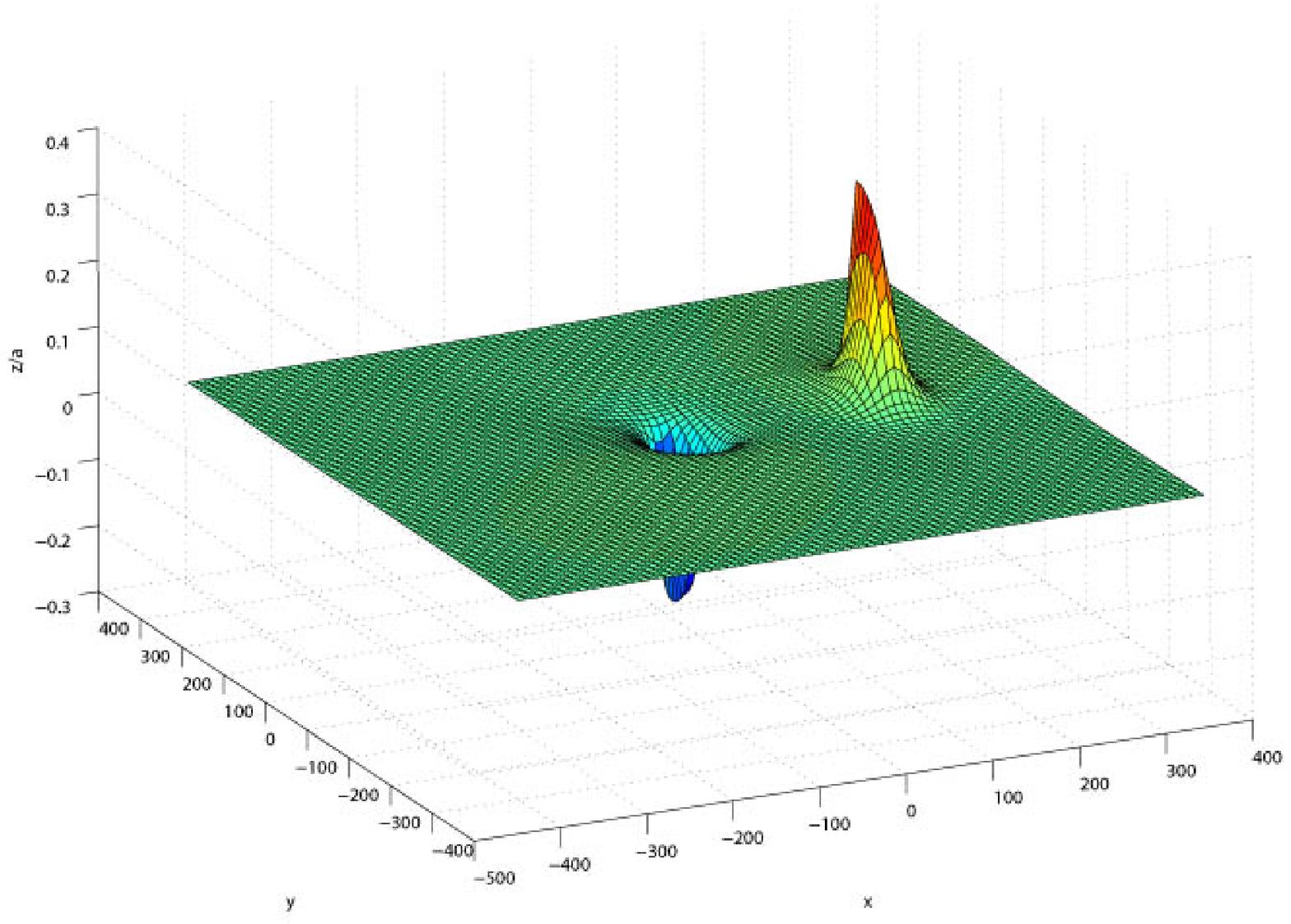}\\
  \caption{Dimensionless free-surface deformation $z/a$ after strike-slip fault. Here $a$
is $|\Dv|$ (30 m in the present application).}
  \label{fig:strike}
\end{figure}

\begin{figure}
  \includegraphics[width=\linewidth]{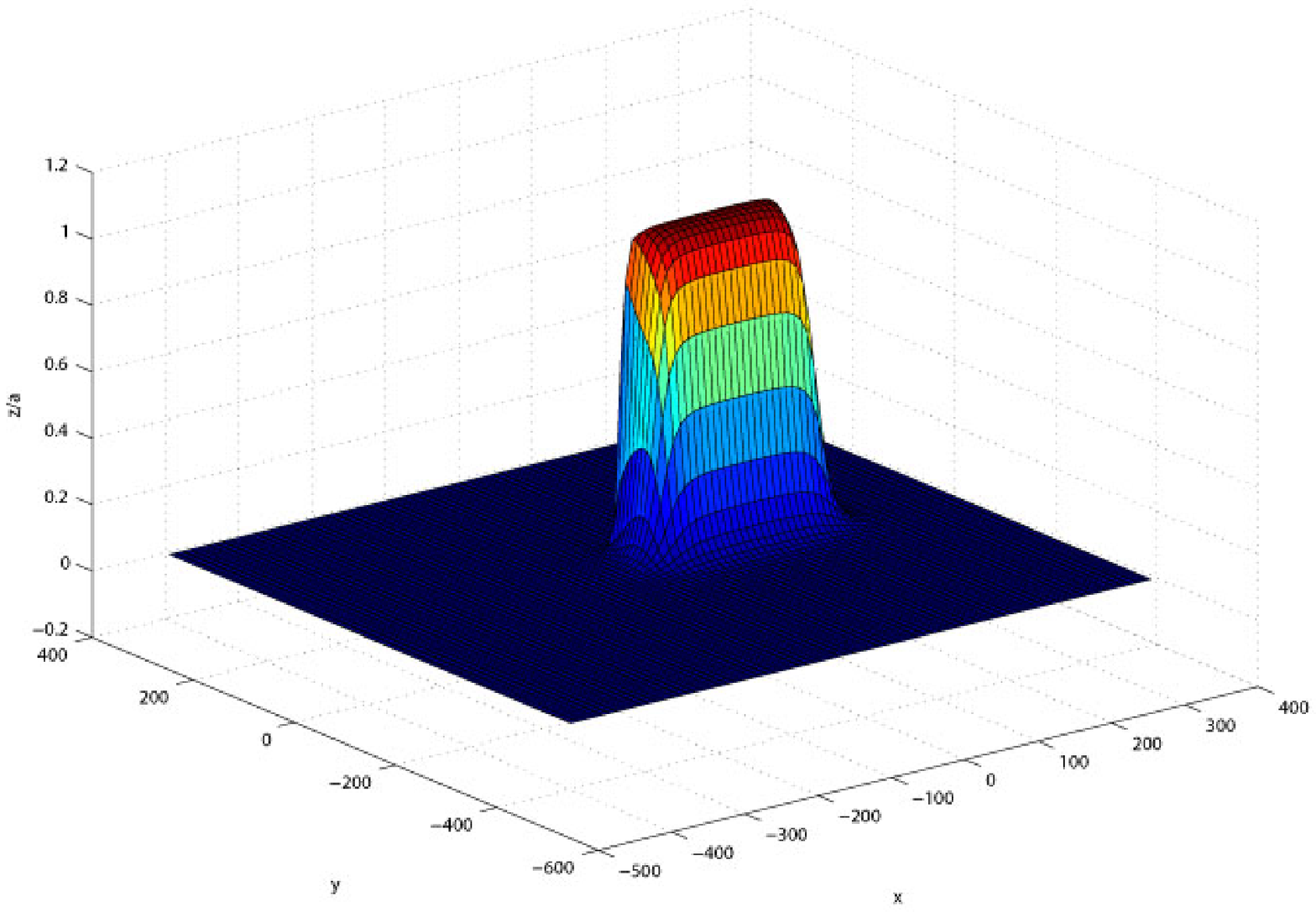}\\
  \caption{Dimensionless free-surface deformation $z/a$ after tensile fault. Here
$\Dv=(0,0,U_3)$ and $a=U_3$.}
  \label{fig:tensile}
\end{figure}

Figures \ref{fig:dip}, \ref{fig:strike}, and \ref{fig:tensile}
show the free-surface deformation after three elementary
dislocations. The values of the parameters are given in Table \ref{parset}.
\begin{table}
\begin{center}
\begin{tabular}{lc}
  \hline
{\it parameter} & {\it value} \\
\hline
  Dip angle $\delta$ & $13^\circ$ \\
  Fault depth $d$, km & 25 \\
  Fault length $L$, km & 220 \\
  Fault width $W$, km & 90 \\
  $U_i$, m & 30 \\
   Young modulus $E$, GPa & 9.5 \\
  Poisson's ratio $\nu$ & 0.23 \\
\hline
\end{tabular}
\end{center}
\caption[]{Parameter set used in Figures \ref{fig:dip}, \ref{fig:strike}, and \ref{fig:tensile}.}
\label{parset}
\end{table}

The traditional approach for hydrodynamic modelers is indeed to use elastic models similar
to the model just described with the seismic parameters as input to evaluate details of the
seafloor deformation. Then this deformation is translated to the initial condition of the
evolution problem described in the next section. A few authors have solved the linearized water
wave equations in the presence of a moving bottom \cite{Hammack,Todo}.

\section{Propagation of tsunamis}

The problem of tsunami propagation is a special case of the general water-wave problem.
The study of water waves relies on several common assumptions. Some are obvious while some 
others are questionable under certain circumstances. The water is assumed to be 
incompressible.
Dissipation is not often included. However there are three main sources of dissipation for water waves:
bottom friction, surface dissipation and body dissipation. For tsunamis, bottom friction is
the most important one, especially in the later stages, and is sometimes included in the computations 
in an ad-hoc way. In most theoretical analyses, it is not included. 
 
A brief description of the common mathematical model used to study water waves follows.
The horizontal coordinates are denoted by $x$ and $y$, and the vertical coordinate by $z$. 
The horizontal gradient is denoted by
$$ \nabla := \left( \frac{\partial}{\partial x}, \frac{\partial}{\partial y} \right). $$
The horizontal velocity is denoted by
$$ \uv(x,y,z,t) = (u,v) $$
and the vertical velocity by $w(x,y,z,t)$.
The three-dimensional flow of an inviscid and incompressible fluid 
is governed by the conservation of mass 
\begin{equation}\label{continuity}
     \nabla \cdot \uv  + \frac{\partial w}{\partial z} = 0 
\end{equation}
and by the conservation of momentum 
\begin{equation}\label{momentum}
    \rho\bDDt{\uv} = -\nabla p, \quad
\rho\bDDt{w} = -\rho g - \frac{\partial p}{\partial z}.
\end{equation}
In (\ref{momentum}), 
$\rho$ is the density of water (assumed to be constant throughout
the fluid domain), $g$ is the acceleration 
due to gravity and $p(x,y,z,t)$ the pressure field.

The assumption that the flow is irrotational is commonly made to analyze surface waves.  
Then there exists a scalar function $\phi(x,y,z,t)$ (the velocity potential) such that 
$$\uv = \nabla \phi, \quad w = \frac{\partial\phi}{\partial z}.$$ 
The continuity equation (\ref{continuity}) becomes
\begin{equation}\label{laplace}
    \nabla^2 \phi + \frac{\partial^2\phi}{\partial z^2} = 0 \,.   
\end{equation}
The equation of momentum conservation (\ref{momentum}) 
can be integrated into Bernoulli's equation
\begin{equation}\label{bernoulli}
    \frac{\partial \phi}{\partial t}+\frac{1}{2}|\nabla \phi|^2
+\frac{1}{2}\left(\frac{\partial \phi}{\partial z}\right)^2+gz + \frac{p-p_0}{\rho} = 0 \,,  
\end{equation}
which is valid everywhere in the fluid. The constant $p_0$ is a pressure of reference, for example the 
atmospheric pressure. The effects of surface tension are not important for tsunami propagation.

\subsection{Classical formulation}

The surface wave problem consists in solving Laplace's equation (\ref{laplace}) in a domain $\Omega(t)$ 
bounded above by a moving free surface (the interface between air and water)
and below by a fixed solid boundary (the bottom).\footnote{The surface wave problem can be easily extended
to the case of a moving bottom. This extension may be needed to model tsunami generation if the
bottom deformation is relatively slow.}
The free surface is represented by $F(x,y,z,t) :=
\eta(x,y,t)-z = 0$.  The shape of the bottom is given by 
$z=-h(x,y)$. The main driving force is gravity.

The free surface must be found as part of the solution. Two boundary conditions are required.  
The first one is the kinematic condition. It can be stated as $\rD F/\rD t = 0$ (the material derivative of $F$
vanishes), which leads to
\begin{equation}\label{kbc}
    \eta_t + \nabla \phi \cdot \nabla \eta - \phi_z = 0 \quad \mbox{at} \;\; z = \eta(x,y,t)\,.
\end{equation} 
    The second boundary condition is the dynamic condition which states that
    the normal stresses must be in balance at the free surface.  
The normal stress at the free surface is given by the difference in pressure.
Bernoulli's equation (\ref{bernoulli}) evaluated on the free surface $z=\eta$ gives
\begin{equation}\label{bdc}
    \phi_t+\fr|\nabla \phi|^2 + \fr \phi_z^2 +g\eta = 0 
\quad \mbox{at} \;\; z = \eta(x,y,t)\,. 
\end{equation}
    
Finally, the boundary condition at the bottom is
\begin{equation}\label{bottom}
    \nabla \phi \cdot \nabla h + \phi_z = 0 \quad \mbox{at} \;\; z = -h(x,y)\,. 
\end{equation}

To summarize, the goal is to solve the set of equations (\ref{laplace}),
(\ref{kbc}), (\ref{bdc}) and (\ref{bottom})
for $\eta(x,y,t)$ and $\phi(x,y,z,t)$.
When the initial value problem is integrated, the fields
$\eta(x,y,0)$ and $\phi(x,y,z,0)$ must be specified at $t=0$.
The conservation of momentum equation (\ref{momentum}) is not 
required in the solution procedure; it is used 
\emph{a posteriori} to find the pressure $p$ once $\eta$ and 
$\phi$ have been found.

In the following subsections, we will consider various approximations of the full 
water-wave equations. One is the system of
Boussinesq equations, that retains nonlinearity and dispersion up to a certain
order. Another one is the system of nonlinear shallow-water equations that
retains nonlinearity but no dispersion. The simplest one is the system of linear
shallow-water equations. The concept of shallow water is based on the smallness
of the ratio between water depth and wave length. In the case of tsunamis propagating
on the surface of deep oceans, one can consider that shallow-water theory is 
appropriate because the water depth (typically several kilometers) is much smaller
than the wave length (typically several hundred kilometers). 

\subsection{Dimensionless formulation}

The derivation of shallow-water type equations is a classical topic.  
Two dimensionless numbers, which are supposed to be small, are introduced:
\begin{equation}
\alpha = \frac{a}{d} \ll 1, \quad \beta=\frac{d^2}{\ell^2} \ll 1,
\end{equation}
where $d$ is a typical water depth, $a$ a typical wave amplitude and
$\ell$ a typical wavelength. The assumptions on the smallness of these two numbers are satisfied
for the Indian Ocean tsunami. Indeed the satellite altimetry observations of the tsunami waves
obtained by two satellites that passed over the Indian Ocean a couple of hours after the
rupture occurred give an amplitude $a$ of roughly 60 cm in the open ocean. The typical 
wavelength estimated from the width of the 
segments that experienced slip is between 160 and 240 km \cite{Lay}. The water depth ranges from 4 km
towards the west of the rupture to 1 km towards the east. These values give the following ranges for the 
two dimensionless numbers:
\begin{equation}
1.5 \times 10^{-4} < \alpha < 6 \times 10^{-4}, \quad 1.7 \times 10^{-5} < \beta < 6.25 \times 10^{-4}.
\end{equation}
The equations are more transparent when written in dimensionless variables. The new independent variables
are 
\begin{equation}
x=\ell \tilde x, \quad y=\ell \tilde y, \quad z=d\tilde z, 
\quad t=\ell \tilde t/c_0, 
\end{equation}
where $c_0=\sqrt{gd}$, the famous speed of propagation of tsunamis in the open ocean ranging from
356 km/h for a 1 km water depth to 712 km/h for a 4 km water depth. The new dependent variables are
\begin{equation}
\eta=a\tilde\eta, \quad h = d \tilde h, \quad \phi=ga\ell\tilde\phi/c_0. 
\end{equation}
In dimensionless form, and after dropping the tildes, the equations become
\begin{eqnarray}
\beta \nabla^2 \phi + \phi_{zz} & = & 0, \\
\beta \nabla \phi \cdot \nabla h + \phi_z & = & 0 \quad \mbox{at} \;\; z = -h(x,y), \\
\label{kin} \beta \eta_t + \alpha\beta \nabla \phi \cdot \nabla \eta & = & \phi_z \quad \mbox{at} 
\;\; z = \alpha\eta(x,y,t), \\ 
\label{dyn} \beta\phi_t+\fr\alpha\beta|\nabla \phi|^2 + \fr\alpha \phi_z^2 + \beta\eta & = & 0 
\quad \mbox{at} \;\; z = \alpha\eta(x,y,t). 
\end{eqnarray}
So far, no approximation has been made. In particular, we have not used the fact that the numbers
$\alpha$ and $\beta$ are small. 

\subsection{Shallow-water equations}

When $\beta$ is small, the water is considered to be shallow. The linearized theory of water
waves is recovered by letting $\alpha$ go to zero. For the shallow water-wave theory, one assumes
that $\beta$ is small and expand $\phi$ in terms of $\beta$:
$$ \phi = \phi_0 + \beta \phi_1 + \beta^2 \phi_2 + \cdots. $$
This expansion is substituted into the governing equation and the boundary conditions. The lowest-order
term in Laplace's equation is
\begin{equation}
\phi_{0zz} = 0.
\end{equation}
The boundary conditions imply that $\phi_0=\phi_0(x,y,t)$. Thus the vertical velocity component
is zero and the horizontal velocity components are independent of the vertical coordinate $z$
at lowest order. Let $\phi_{0x}=u(x,y,t)$ and $\phi_{0y}=v(x,y,t)$. Assume now for simplicity
that the water depth is constant ($h=1$). Solving Laplace's equation and taking
into account the bottom kinematic condition yields the following expressions for $\phi_1$ and $\phi_2$:
\begin{eqnarray}
\phi_1(x,y,z,t) & = & -\fr (1+z)^2(u_x+v_y), \\
\phi_2(x,y,z,t) & = & \onetf (1+z)^4[(\nabla^2 u)_x + (\nabla^2 v)_y].
\end{eqnarray}
The next step consists in retaining terms of requested order in the free-surface boundary conditions. 
Powers of $\alpha$ will appear when expanding in Taylor series the free-surface conditions around
$z=0$. For example, if one keeps terms of order $\alpha\beta$ and $\beta^2$ in the dynamic boundary 
condition (\ref{dyn}) and in the kinematic boundary condition (\ref{kin}), one obtains
\begin{eqnarray}
\label{dyn1} \beta\phi_{0t} - \fr \beta^2 (u_{tx}+v_{ty}) + \beta\eta + \fr \alpha\beta (u^2+v^2) & = & 0, \\
\label{kin1} \beta [\eta_t + \alpha(u\eta_x+v\eta_y) + (1+\alpha\eta)(u_x+v_y)] & = & \ones \beta^2
[(\nabla^2 u)_x + (\nabla^2 v)_y].
\end{eqnarray}
Differentiating (\ref{dyn1}) first with respect to $x$ and then to respect to $y$ gives a set of two
equations:
\begin{eqnarray}
\label{dynx} u_t + \alpha (uu_x+vv_x) + \eta_x - \fr \beta (u_{txx}+v_{txy}) & = & 0, \\
\label{dyny} v_t + \alpha (uu_y+vv_y) + \eta_y - \fr \beta (u_{txy}+v_{tyy}) & = & 0.
\end{eqnarray}
The kinematic condition (\ref{kin1}) can be rewritten as
\begin{equation}
\label{kin2} \eta_t + [u(1+\alpha\eta)]_x + [v(1+\alpha\eta)]_y = \ones \beta
[(\nabla^2 u)_x + (\nabla^2 v)_y].
\end{equation}
Equations (\ref{dynx})--(\ref{kin2}) contain in fact various shallow-water models. The so-called
fundamental shallow-water equations are obtained by neglecting the terms of order $\beta$:
\begin{eqnarray}
\label{sw1} u_t + \alpha (uu_x+vu_y) + \eta_x & = & 0, \\
\label{sw2} v_t + \alpha (uv_x+vv_y) + \eta_y & = & 0, \\
\label{sw3} \eta_t + [u(1+\alpha\eta)]_x + [v(1+\alpha\eta)]_y & = & 0.
\end{eqnarray}
Recall that we assumed $h$ to be constant for the derivation. Going back to an arbitrary 
water depth and to dimensional variables, the system of nonlinear shallow water equations reads
\begin{eqnarray}
\label{cg1} u_t + uu_x+vu_y + g\eta_x & = & 0, \\
\label{cg2} v_t + uv_x+vv_y + g\eta_y & = & 0, \\
\label{cg3} \eta_t + [u(h+\eta)]_x + [v(h+\eta)]_y & = & 0.
\end{eqnarray}
This system of equations has been used for example by Titov and Synolakis for the numerical
computation of tidal wave run-up \cite{TS}. Note that this model does not include any bottom
friction terms. To solve the problem of tsunami generation caused by bottom displacement, 
the motion of the seafloor obtained from seismological models \cite{Okada85} and described in
Section 3 can be prescribed during 
a time $t_0$. Usually $t_0$ is assumed to be small, so that the bottom displacement is considered
as an instantaneous vertical displacement. This assumption may not be appropriate for slow events. 

The satellite altimetry observations of the Indian Ocean tsunami clearly show dispersive effects.
The question of dispersive effects in tsunamis is open. Most propagation codes ignore dispersion.
A few propagation codes that include dispersion have been developed \cite{DGK}. A well-known code is FUNWAVE,
developed at the University of Delaware over the past ten years \cite{Fun}. Dispersive shallow 
water-wave models are presented next.

\subsection{Boussinesq equations}

An additional dimensionless number, sometimes called the Stokes number, is introduced:
\begin{equation}
S=\frac{\alpha}{\beta} \approx 1.
\end{equation}
For the Indian Ocean tsunami, one finds 
\begin{equation}
0.24 < S < 46.
\end{equation}
Therefore the additional assumption that $S\approx 1$ may be realistic. 

In this subsection, we provide the guidelines to derive Boussinesq-type systems of equations
\cite{BCS}. Of course,
the variation of bathymetry is essential for the propagation of tsunamis, but for the
derivation the water depth will be assumed to be constant. 
Some notation is introduced. The potential evaluated along 
the free surface is denoted by $\Phi(x,y,t):=\phi(x,y,\eta,t)$. 
The derivatives of the velocity potential evaluated on the free surface are denoted by 
$\Phi_{(*)} (x,y,t):=\phi_* (x,y,\eta,t),$ where the star stands for $x$, $y$, $z$ or $t$.  
Consequently, $\Phi_*$ (defined for $* \neq z$) and $\Phi_{(*)}$ have different meanings.  
They are however related since
$$ \Phi_* = \Phi_{(*)} + \Phi_{(z)}\eta_* \,.$$
The vertical velocity at the free surface is denoted by $W(x,y,t):=\phi_z(x,y,\eta,t)$.

The boundary conditions on the free surface (\ref{kbc}) and (\ref{bdc}) become
\begin{eqnarray}
\eta_t + \nabla\Phi \cdot \nabla\eta - W(1+\nabla\eta \cdot \nabla\eta) & = & 0, \\
\Phi_t + g\eta + \fr|\nabla\Phi|^2 - \fr W^2(1+\nabla\eta \cdot \nabla\eta) & = & 0.
\end{eqnarray}
These two nonlinear equations provide time-stepping for $\eta$ and $\Phi$.
In addition, Laplace's equation as well as the kinematic condition on the bottom must be
satisfied. In order to relate the free-surface variables with the bottom variables, one
must solve Laplace's equation in the whole water column. In Boussinesq-type models,
the velocity potential is represented as a formal expansion,
\begin{equation}
\phi(x,y,z,t) = \sum_{n=0}^\infty \phi^{(n)}(x,y,t)\, z^n. 
\end{equation}
Here the expansion is about $z=0$, which is the location of the free surface at rest. Demanding that 
$\phi$ formally satisfy Laplace's equation
leads to a recurrence relation between $\phi^{(n)}$ and $\phi^{(n+2)}$. Let $\phi_o$ denote
the velocity potential at $z=0$, $\uv_o$ the horizontal velocity at $z=0$, and $w_o$ the vertical 
velocity at $z=0$. Note that $\phi_o$
and $w_o$ are nothing else than $\phi^{(0)}$ and $\phi^{(1)}$. The potential $\phi$ can be 
expressed in terms of $\phi_o$ and $w_o$ only. Finally, one obtains the velocity field in the whole water
column $(-h \le z \le \eta)$ \cite{Madsen03}:
\begin{eqnarray}
\label{uc} \uv(x,y,z,t) & = & \cos(z\nabla)\uv_o + \sin(z\nabla) w_o, \\
\label{wc} w(x,y,z,t) & = & \cos(z\nabla) w_o - \sin(z\nabla)\uv_o.
\end{eqnarray}
Here the cosine and sine operators are infinite Taylor series operators defined by
$$ \cos(z\nabla) = \sum_{n=0}^\infty (-1)^n \frac{z^{2n}}{(2n)!}\nabla^{2n}, \quad 
\sin(z\nabla) = \sum_{n=0}^\infty (-1)^n \frac{z^{2n+1}}{(2n+1)!}\nabla^{2n+1}. $$

Then one can substitute the representation (\ref{uc})-(\ref{wc}) into the kinematic bottom condition
and use successive approximations to obtain an explicit recursive expression for $w_o$ in terms
of $\uv_o$ to infinite order in $h\nabla$.

A wide variety of Boussinesq systems can been derived \cite{Madsen03}. One can generalize the
expansions to an arbitrary $z-$level, instead of the $z=0$ level. The Taylor series for the cosine 
and sine operators can
be truncated, Pad\'e approximants can be used in operators at $z=-h$ and/or at $z=0$. 

The classical Boussinesq equations are more transparent when written in the dimensionless variables
used in the previous subsection. We further assume that $h$ is constant,
drop the tildes, and write the equations for one spatial dimension ($x$). Performing the expansion 
about $z=0$ leads to the vanishing of
the odd terms in the velocity potential. Substituting the expression 
for $\phi$ into the free-surface boundary conditions evaluated at $z=1+\alpha\eta(x,t)$ leads to
two equations in $\eta$ and $\phi_o$ with terms of various order in $\alpha$ and $\beta$. The small
parameters $\alpha$ and $\beta$ are of the same order, while $\eta$ and $\phi_o$ as well as their partial
derivatives are of order one. 

\subsection{Classical Boussinesq equations}

The classical Boussinesq equations are obtained by keeping all terms that are at most linear in $\alpha$ or 
$\beta$. In the derivation of the fundamental nonlinear shallow-water equations (\ref{sw1})--(\ref{sw3}), 
the terms in $\beta$ were
neglected. It is therefore implicitly assumed that the Stokes number is large. Since the cube of the
water depth appears in the denominator of the Stokes number ($S=\alpha/\beta=a\ell^2/d^3$), it means 
that the Stokes number is 64 times larger in a 1 km depth than in a 4 km depth! Based on these arguments, 
dispersion is more important to the west of the rupture. Considering the Stokes number to be of order one 
leads to the following system in dimensional form\footnote{Equations (\ref{cb1}) and (\ref{cb2}) could
have been obtained from equations (\ref{dynx}) and (\ref{kin2}).}:
\begin{eqnarray}
\label{cb1} u_t + uu_x + g\eta_x - \fr h^2 u_{txx} & = & 0, \\
\label{cb2} \eta_t + [u(h+\eta)]_x - \ones h^3 u_{xxx} & = & 0.
\end{eqnarray}
The classical Boussinesq equations are in fact slightly different. They are obtained by replacing
$u$ with the depth averaged velocity 
$$ \frac{1}{h} \int_{-h}^\eta u\,dz. $$
They read
\begin{eqnarray}
u_t + uu_x + g\eta_x - \onet h^2 u_{txx} & = & 0, \\
\eta_t + [u(h+\eta)]_x & = & 0.
\end{eqnarray}
A number of variants of the classical Boussinesq system were studied by Bona et al., who in 
particular showed that depending on the modeling of dispersion the linearization about the
rest state may or may not be well-posed \cite{BCS}.

\subsection{Korteweg--de Vries equation}

The previous system allows the propagation of waves in both the positive and negative $x-$directions. Seeking
solutions travelling in only one direction, for example the positive $x-$direction, leads to a
single equation for $\eta$, the Korteweg--de Vries equation:
\begin{equation}
\eta_t + c_0\left(1+\frac{3\eta}{2d}\right)\eta_x + \frac{1}{6} c_0d^2 \eta_{xxx} = 0,
\end{equation}
where $d$ is the water depth.
It admits solitary wave solutions travelling at speed $V$ in the form
\[ \eta(x,t) = a \, \sech^2 \left(\sqrt{\frac{3a}{4d^3}}(x-Vt)\right), \quad \mbox{with} \;\;
V=c_0\left(1+\frac{a}{2d}\right). \]
The solitary wave solutions of the Korteweg--de Vries equation are of elevation ($a>0$) and 
travel faster than $c_0$. Their speed increases with amplitude. Note that a natural length scale 
appears: 
\[ \ell = \sqrt{\frac{4d^3}{3a}}. \]
For the Indian Ocean tsunami, it gives roughly $\ell = 377$ km. It is of the order of magnitude
of the wavelength estimated from the width of the segments that experienced slip. 

\section{Energy of a tsunami}

The energy of the earthquake is measured via the strain energy released by the faulting. The part 
of the energy transmitted to the tsunami wave is less than one percent \cite{Lay}. They estimate
the tsunami energy to be $4.2 \times 10^{15}$ J. They do not give details on how they obtained 
this estimate. However, a simple calculation based on considering the tsunami as a soliton
$$ \eta(x) = a \, \sech^2 \left(\frac{x}{\ell}\right), \quad u(x)= \alpha c_0 \,
\sech^2 \left(\frac{x}{\ell}\right), $$
gives for the energy
$$ E = \frac{1}{\sqrt 3} \alpha^{3/2} \rho d^2 (c_0^2+gd) \int_{-\infty}^{\infty} \sech^4 x \, dx +
O(\alpha^2). $$
The value for the integral is $4/3$. The numerical estimate for $E$ is close to that of Lay et al. (2005). 
Incidently, at this level of approximation, there is equipartition between kinetic and potential
energy. It is also important to point out that a tsunami being a shallow water wave, the whole 
water column is moving as the wave propagates. For the parameter values used so far, the maximum horizontal
current is 3 cm/s. However, as the water depth decreases, the current increases and becomes important
when the depth becomes less than 500 m. Additional properties of solitary waves can be found for example
in \cite{LH}.

\section{Tsunami run-up}

The last phase of a tsunami is its run-up and inundation. Although in some
cases it may be important to consider the coupling between fluid and structures, we restrict
ourselves to the description of the fluid flow. 
The problem of waves climbing a beach is a classical one \cite{CG58}. The transformations used
by Carrier and Greenspan are still used nowadays. The basis of their analysis is the one-dimensional
counterpart of the system (\ref{cg1})--(\ref{cg3}). In addition, they assume the depth to be of
uniform slope: $h = -x\tan\theta$. Introduce the following dimensionless quantities, where $\ell$ 
is a characteristic length\footnote{In fact there is no obvious characteristic length in this idealized
problem. Some authors simply say at this point that $\ell$ is specific to the problem under
consideration.}:
\[ x=\ell\tilde x, \;\; \eta=\ell\tilde\eta, \;\; u=\sqrt{g\ell}\,\tilde u,
\;\; t=\sqrt{\ell/g}\,\tilde t, \;\; c^2=(h+\eta)/\ell. \]
After dropping the tildes, the dimensionless system of equations (\ref{cg1})-(\ref{cg3}) becomes
\begin{eqnarray*}
u_t + uu_x + \eta_x & = & 0, \\
\eta_t + [u(-x\tan\theta+\eta)]_x & = & 0.
\end{eqnarray*}
In terms of the variable $c$, these equations become
\begin{eqnarray*}
u_t + uu_x + 2cc_x + \tan\theta & = & 0, \\
2c_t + cu_x + 2uc_x & = & 0.
\end{eqnarray*}
The equations written in characteristic form are
\begin{eqnarray*}
\left[\frac{\partial}{\partial t} + (u+c)\frac{\partial}{\partial x}\right](u+2c+t\tan\theta) & = & 0, \\
\left[\frac{\partial}{\partial t} + (u-c)\frac{\partial}{\partial x}\right](u-2c+t\tan\theta) & = & 0.
\end{eqnarray*}
The characteristic curves $C^+$ and $C^-$ as well as the Riemann invariants are 
\begin{eqnarray*}
C^+ & : & \frac{dx}{dt} = u+c, \quad u+2c+t\tan\theta = r, \\
C^- & : & \frac{dx}{dt} = u-c, \quad u-2c+t\tan\theta = s.
\end{eqnarray*}
Next one can rewrite the hyperbolic equations in terms of the new variables $\lambda$ and $\sigma$
defined as follows:
\begin{eqnarray*}
\frac{\lambda}{2} & = & \frac{1}{2}(r+s) = u + t\tan\theta, \\
\frac{\sigma}{4} & = & \frac{1}{4}(r-s) = c.
\end{eqnarray*}
One obtains
\begin{eqnarray*}
x_s - \left[\frac{1}{4}(3r+s)-t\tan\theta\right]t_s & = & 0, \\
x_r - \left[\frac{1}{4}(r+3s)-t\tan\theta\right]t_r & = & 0.
\end{eqnarray*}
The elimination of $x$ results in the {\it linear} second-order equation for $t$
\begin{equation}
\label{eq-t} \sigma(t_{\lambda\lambda}-t_{\sigma\sigma})-3t_\sigma = 0. 
\end{equation}
Since $u + t\tan\theta = \lambda/2$, $u$ must also satisfy (\ref{eq-t}). 
Introducing the potential $\phi(\sigma,\lambda)$ such that
\[ u = \frac{\phi_\sigma}{\sigma}, \]
one obtains the equation
\[ (\sigma\phi_\sigma)_\sigma - \sigma \phi_{\lambda\lambda} = 0 \]
after integrating once. Two major simplifications have been obtained. The nonlinear
set of equations have been reduced to a linear equation for $u$ or $\phi$ and the free
boundary is now the fixed line $\sigma=0$ in the $(\sigma,\lambda)-$plane. The free 
boundary is the instantaneous shoreline $c=0$, which moves as a wave climbs a beach.

The above formulation has been used by several authors to study the run-up of various
types of waves on sloping beaches \cite{TS94,CWY,Tinti05}. For example, it has been shown
that leading depression $N$-waves run-up higher than leading elevation $N$-waves, suggesting
that perhaps the solitary wave model may not be adequate for predicting an upper limit for
the run-up of near-shore generated tsunamis. 

There is a rule of thumb that says that the run-up does not usually exceed twice the fault slip.
Since run-ups of 30 meters were observed in Sumatra during the Boxing Day tsunami, the slip
might have been of 15 meters or even more. 

Analytical models are useful, especially to perform parametric studies. However, the breaking
of tsunami waves as well as the subsequent floodings must be studied numerically. 
The most natural methods that can be used are
the free surface capturing methods based on a finite volume discretisation, such as the Volume Of Fluid 
(VOF) or the Level Set methods, and the family of Smoothed Particle Hydrodynamics methods (SPH), 
applied to free-surface flow problems \cite{Monaghan,SPH1,SPH2}. Such methods allow a study
of flood wave dynamics, of wave breaking on the land behind beaches, and of the flow over rising ground 
with and without the presence of 
obstacles. This task is an essential part of tsunami modelling, since it allows the determination of 
the level of risk due to major flooding, the prediction of the resulting water levels in the flooded 
areas, the determination of security zones.  It also provides some help in the conception and validation 
of protection systems in the most exposed areas.  

\section{Direction for future research}

A useful direction for future research in the dynamics of tsunami waves is the three-dimensional (3D)
simulation of tsunami breaking 
along a coast.  For this purpose, different validation steps are necessary.
First more simulations of a two-dimensional (2D) tsunami interacting with a sloping beach
ought to be performed.  Then these simulations should be extended to the case of a 2D tsunami 
interacting with a sloping beach in the presence of obstacles. An important output of these
computations will be the hydrodynamic loading on obstacles.  The nonlinear inelastic behaviour of the 
obstacles may be accounted for using damage or plasticity models. The development of Boussinesq
type models coupled with structure interactions is also a promising task. Finally there is a need for
3D numerical simulations of a 
tsunami interacting with a beach of complex bathymetry, with or without obstacles.  
These simulations will hopefully demonstrate the usefulness 
of numerical simulations for the definition of protecting devices or security zones.  
An important challenge in that respect is to make the numerical methods capable of handling 
interaction problems involving different scales: the fine scale needed for representing 
the damage of a flexible obstacle and a coarse scale needed to quantify the tsunami propagation.

\bibliography{tsunami}
\bibliographystyle{plainnat}

\end{document}